\RequirePackage{luatex85}
\documentclass{jfm}

\usepackage[usenames,dvipsnames]{xcolor}
\usepackage{graphicx}
\usepackage{epstopdf, epsfig}
\usepackage{amsmath}
\usepackage{amssymb}
\usepackage{amsfonts}
\usepackage{natbib}
\usepackage{psfrag}
\usepackage{wasysym}
\usepackage{float}
\usepackage{color}
\usepackage[normalem]{ulem} 
\usepackage{textcomp}
\usepackage{multirow,bm}
\usepackage{makecell}
\usepackage{ulem}
\usepackage{epstopdf}

\newcommand{\pd}[2]{\frac{\partial #1}{\partial #2}} 
\newcommand{\pdd}[2]{\frac{\partial^2 #1}{\partial #2^2}} 

\title{Regimes of Soft Lubrication}

\author{Martin H. Essink\aff{1}\corresp{\email{m.h.essink@utwente.nl}}, Anupam Pandey\aff{1,2}, Stefan Karpitschka\aff{3}, Cornelis H. Venner\aff{4} and Jacco H. Snoeijer\aff{1}}
\affiliation{
	\aff{1}Physics of Fluids Group, Faculty of Science and Technology, Mesa+ Institute, University of Twente, 7500 AE Enschede, The Netherlands.\\
	\aff{2}Department of Biological and Environmental Engineering, Cornell University, Ithaca, NY 14853, USA.\\
	\aff{3}Max Planck Institute for Dynamics and Self-Organization (MPIDS), 37077 G\"ottingen, Germany.\\
	\aff{4}Faculty of Engineering Technology, Engineering Fluid Dynamics, University of Twente, P.O. Box 217, 7500 AE Enschede, The Netherlands.}

\shorttitle{Regimes of Soft Lubrication}
\shortauthor{Essink, Pandey, Karpitschka, Venner \& Snoeijer}

\begin{document}
	
\maketitle

\begin{abstract}
	Elastohydrodynamic lubrication, or simply soft lubrication, refers to the motion of deformable objects near a boundary lubricated by a fluid, and is one of the key physical mechanisms to minimise friction and wear in natural and engineered systems. Hence it is of particular interest to relate the thickness of the lubricant layer to the entrainment (sliding/rolling) velocity, the mechanical loading exerted onto the contacting elements, and properties of the elastic boundary. In this work we provide an overview of the various regimes of soft lubrication for two-dimensional cylinders in lubricated contact with compliant walls. We discuss the limits of small and large entrainment velocity, which is equivalent to large and small elastic deformations, as the cylinder moves near thick or thin elastic layers. The analysis focusses on thin elastic coatings, both compressible and incompressible, for which analytical scaling laws are not yet available in the regime of large  deformations.	By analysing the elastohydrodynamic boundary layers that appear at the edge of the contact, we establish the missing scaling laws -- including prefactors. As such, we offer a rather complete overview of  physically relevant limits of soft lubrication.
\end{abstract}


\section{Introduction}

The introduction of a viscous liquid within the narrow gap between two moving bodies prevents solid-solid contact, reduces friction, minimizes wear, and allows precise motion control; a process known as lubrication. Lubrication has played a pivotal role in major engineering milestones, from revolutionising transportation of heavy objects on wooden sledges in ancient Egypt~\citep[]{dowson} to the efficient and reliable functioning of bearings in applications on many scales ranging up to the huge  rolling  bearings in todays wind turbines. The underlying hydrodynamic framework of fluid flow in thin gaps is in fact relevant to a range of phenomena across different length scales: motion of drops/bubbles in microfluidic devices~\citep[]{Bretherton1961,Byun2013a}, moving blood cells in capillaries~\citep[]{Fitz1969, Secomb1986}, and biomechanics of synovial joints~\citep[]{Hou1992} to name a few. However, in most of the engineering applications as well as natural systems, one or both of the moving boundaries are deformable. As a result, lubrication flow is coupled to elasticity through the non-local soft lubrication equations.

Since the realization of the essential coupling between deformation and flow for film formation, the study of elasto-hydrodynamic lubrication in engineering (Tribology) has established as a separate field of research. Its importance has only increased with the current need of reduced material and lubricant in view of sustainability demands. More complex time-varying and extreme operating conditions, such as higher loads and higher temperatures, lead to increased deformation and thinner films which calls for better understanding and improved prediction capability for engineering and design. Only in very specific asymptotic cases analytic solutions could be derived, and numerical solution is often needed. Pioneering work was done by Dowson and Higginson \citep[]{DH1959,DH1966} and \citet[]{HD1977}, where the complexity of the interaction between deformation and flow led to  numerical stability problems at large deformations. Faster computers and improved numerical algorithms \citep[]{hamrock2004fundamentals, venner2000multi} allowed the study of realistic problems related to roller bearings, gears, cam-followers, and seals. Many studies aimed at the derivation of empirical formulas for film thickness prediction under steady conditions from the numerical solutions. Experimental validation was obtained from the further development of optical interferometry based techniques \citep[]{CG1963,CSH1996}. At present, complex cases of contact geometries with time varying loading conditions, roughness moving through the contact, contacts with very limited lubrication supply, grease lubricated contacts, and coated surfaces are considered, see \cite{Wang2019}. In spite of the plethora of problems studied and results presented the fundamental understanding of the physical phenomena and the appropriate scaling laws developed only relatively slowly. A recent overview is given by \citep[]{Greenwood2020}.

A key feature of soft lubrication is the emergence of a non-inertial lift force, which facilitates lubrication by maintaining the gap between the moving surfaces~\citep[]{Sekimoto1993, Skotheim2004}. For micron sized particles flowing in a channel with soft walls, this force causes radial migration of particles~\citep[]{Davies18} and promotes particle sorting in microfluidic devices~\citep[]{GEISLINGER2014161}. A precise measurement of this elastohydrodynamic force can function as a contact-free tool to probe the local rheology of soft materials~\citep[]{leroy2012, wang2015}. For sufficiently soft elastomers, this force is strong enough to sustain the weight of millimeter sized metal cylinders which slide along a deformable wall~\citep[]{Saintyves5847}, and may result in intricate trajectories~\citep[]{Salez2015, Rallabandi17}.

Naturally, it is of fundamental interest to explain how the gap thickness evolves from the coupling between fluid pressure and elastic deformation. The answer depends on details of the physical limit under consideration: Since deformation of a soft layer crucially depends on the geometry, and boundary conditions at the free surface, scaling laws relating the lift force to the lubrication layer are very different in different physical limits. Indeed, recent microfluidic experiments involving polymer brushes, cross linked elastomers, and ionic microgels as soft boundaries have probed elastohydrodynamic interactions in a few different limits, and experimentally found very distinct scaling laws~\cite[]{Davies18, Zhang20, Vialar19}. 

Specifically, the elastic layer can either be thick or thin, the material can be compressible or incompressible, while the elastic deformation can be large or small compared to the lubricant thickness. A well-studied limit is when the soft layer is thin and compressible, so that deformation is proportional to the local pressure, while assuming deformations to be small~\citep[]{Skotheim2005a, Urzay2007, Salez2015}. This allows for a perturbative approach to solve the coupled elastohydrodynamic equations. However, most of the soft materials like elastomers and hydrogels are incompressible, and their response is described by non-local integral equations. Furthermore, if particles are squeezed past a soft layer, a Hertz-like contact forms, where deformation is large as compared to the gap thickness. In this case, the effect of lubrication is primarily dominant within a narrow boundary layer at inlet/outlet regions near the contact edges~\citep[]{Bissett1989, Snoeijer2013}. This lubricated Hertzian limit is not only relevant for classical tribological applications, but also plays a role in soft matter systems such as the slippage of soft microgel pastes~\citep[]{Meeker2004}. Naturally, different scaling laws become relevant in these various limits, but an overview of how these different cases emerge -- and when these are applicable -- is still lacking. 

Here we present a unified overview of different regimes of soft lubrication. Our asymptotic analysis will focus on several important cases that were not considered previously, most notably the lubricated Hertz-like limit for cylinders on thin elastic layers on a rigid substrate. We also explore the transition to the small deformation limit, using numerical simulations. The paper is organised as follows. In section~\ref{sec:framework}, we develop the theoretical and numerical framework and discuss the numerous length scales of the problem. Here we emphasise how the relative magnitudes of these length scales determine the different asymptotic regimes of soft lubrication. In section~\ref{sec:thincomp} we resolve in detail the lubrication on thin compressible layers in the limit of large deformation, and establish the scaling laws through similarity solutions. In the final section~\ref{sec:discuss} this is complemented by the results for thin incompressible layers. Hence, we provide a complete overview of all regimes, summarised in a table in the concluding section, and discuss experimental consequences.  

\begin{figure}
	\centering
	\includegraphics[width=\linewidth]{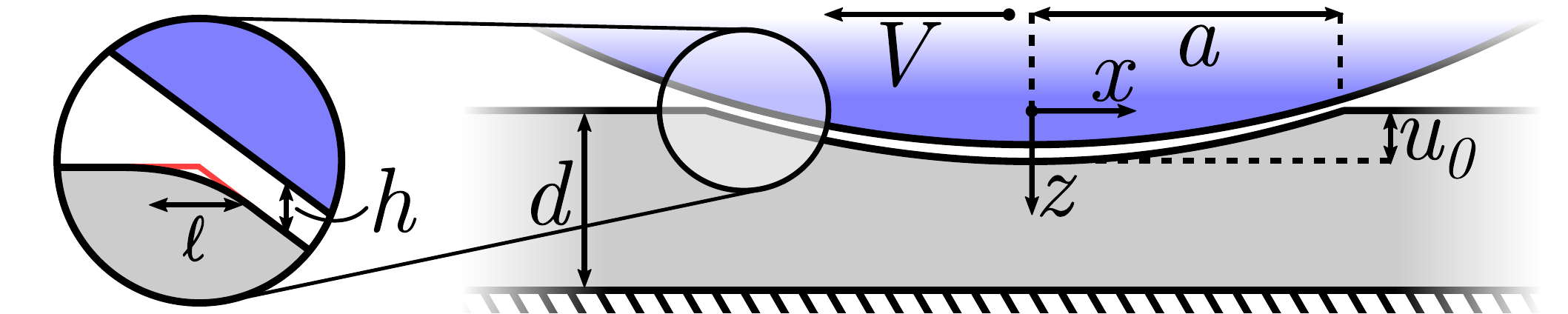}
	\caption{
		Schematic representation of a lubricated contact with a thin, compressible elastic layer, showing the relevant length scales of the problem: elastic layer thickness $d$, contact size $a$, the fluid film thickness $h(x)$ and the elastic deformation $u(x)$ defined in downward direction. One of the edges has been magnified to show the boundary layer width $\ell$, over which lubrication forces smooth out the  elastic deformation.}
	\label{fig:schematic}
\end{figure}

\section{Framework}\label{sec:framework}
\subsection{Problem formulation}

We consider the two-dimensional problem of a rigid cylinder moving at constant velocity parallel to a deformable substrate, as sketched in fig.~\ref{fig:schematic}. Here we consider the translation of the cylinder, but we remark that the analysis is identical for the case that the cylinder is rotating \citep{pandey2016}.  

Importantly, the analysis is organised in terms of the length scales that appear in the problem (fig.~\ref{fig:schematic}). The hierarchy of the length scales is a crucial aspect of the lubrication problem, as the relative sizes of these lengths determine the asymptotic limit. Hence, the organisation of lengths scale will determine the scaling laws that relate the load and velocity to the thickness of the lubrication layer. 

A first aspect of the length scales is that we assume the cylinder radius, $R$, to be much larger than the width of the contact zone, $a$, so that the shape can locally be approximated as a parabola. The lubricant layer thickness, $h(x)$, and the elastic deformation, $u(x)$, are then related by (fig.~\ref{fig:schematic})

\begin{equation} \label{eq:Geometry_equation}
h(x)-\frac{x^2}{2R}+c=u(x).
\end{equation}
Here $c\equiv u(0)-h(0)$ is the indentation of the bottom of the cylinder, measured with respect to the undeformed elastic substrate. Another obvious length scale is the thickness $d$ of the elastic layer. Finally, we anticipate the emergence of a dynamic length scale $\ell$, sketched in the inset of fig.~\ref{fig:schematic}, which acts as a narrow boundary layer at the edges of the contact \citep{Snoeijer2013}. We will find that the relative magnitudes of $a,d,\ell, u$, and the compressibility of the elastic layer (quantified by the Poisson ratio $\nu$), will determine how the lubricant layer $h$ scales with the sliding velocity. We remark that, apart from $d$, none of these lengths are known a priori, but follow from a consistent analysis of the problem. 

Now that the geometry is in place, we can turn to the mechanical equations. We start with the flow of liquid within the narrow gap, which is described by the Stokes equation, $\boldsymbol{\nabla} p = \eta\nabla^2 \boldsymbol{v}$, where $p$ is pressure, $\eta$ is viscosity and $\boldsymbol{v}$ is the velocity field. We consider a reference system where the cylinder is stationary, and the substrate moves with a velocity $V$ to the right. Since both the film thickness and its gradient are small compared to the width of the contact, using no slip boundary conditions the velocity profile within the gap can be written in the lubrication limit,

\begin{equation}
v_x\left(x,z\right)=\frac{1}{2\eta}\pd{p}{x}\Big[(z-u)^2+(z-u)h\Big]+\Big[1+\frac{z-u}{h}\Big]V.
\label{velo}
\end{equation}
Incompressibility of the liquid requires that, at steady-state, the flux $Q=\int^{u}_{u-h}v_x dz$ remains constant. As a result, integration of \eqref{velo} gives the Reynolds equation~\citep[]{Reynolds1886},

\begin{equation} \label{eq:Reynolds_equation}
\pd{p}{x}=6V\eta\frac{h-h^*}{h^3},
\end{equation} where $h^{\ast}=2Q/V$.

Next, we couple the pressure $p(x)$ to the elastic deformation $u(x)$ of the solid. This is most conveniently done by transforming the problem to the Fourier domain ($\tilde{f}(q)=\int_{-\infty}^{\infty}f(x)e^{-iqx}\ dx$; $f(x)=\int_{-\infty}^{\infty}\tilde{f}(q)e^{iqx}\ \frac{dq}{2\pi}$). Namely, the elastic deformation of the substrate in Fourier domain is given by

\begin{equation} \label{eq:Deformation_by_kernel_Fourier}
\tilde{u}(q)={\cal K}(q)\tilde{p}(q),
\end{equation}
where the Green's function ${\cal K}(q)$ reads \citep{hannah1951contact,Wang2019},
\begin{equation} \label{eq:kernel}
{\cal K}(q)=\frac{(1-\nu)}{Gq}\left[\frac{(3-4\nu)\sinh(2qd)-2qd}{(3-4\nu)\cosh(2qd)+2(qd)^2+5-12\nu+8\nu^2}\right].
\end{equation}
In this expression $\nu$ is the Poisson ratio, $G$ is the shear modulus, while $d$ is the layer thickness. For a given pressure, the backward transform to $u(x)$ is to be performed numerically. However, the above kernel can be simplified for various limiting cases, leading to analytical expressions of $u(x)$ in real-space. We distinguish between two limits defined by the quantity $qd$, a ratio of the substrate thickness to the typical wavelength $q$ of the deformation. In the short wavelength limit ($qd\gg1$) the substrate's response locally reduces to that of an elastic half space ${\cal K}(q)\sim|q|^{-1}$. In the opposite limit of large wavelength or thin elastic layer ($qd\ll1$), ${\cal K}(q)$ simplifies to different forms, respectively, for incompressible ($\nu=1/2$) and compressible ($\nu \neq 1/2$) materials. These results along with the corresponding deformations are summarised in table \ref{tbl:elasticity_asymptotes}.

The analysis can be closed by either imposing a vertical load exerted on the cylinder, or equivalently by fixing its vertical position. In our calculations we will impose the total vertical force $L$, expressed in terms of the fluid pressure,

\begin{equation} \label{eq:Lift_integral}
L=\int_{-\infty}^{\infty}p(x)dx.
\end{equation}
The problem is defined by the system of three equations, (\ref{eq:Geometry_equation}, \ref{eq:Reynolds_equation}, \ref{eq:Deformation_by_kernel_Fourier}), which describe the three unknown fields $h(x)$, $p(x)$ and $u(x)$. Furthermore the constraint~\eqref{eq:Lift_integral} enables us to find the vertical position $c$ of the cylinder. Combined with the boundary conditions $p(\pm\infty)=0$ and $u(\pm\infty)=0$, the problem is fully defined. 

\begin{table}
	\centering
		
	\begin{tabular}{lll}
		\hline
		& \textbf{Kernel expansion} & \textbf{Deformation relation} \\ 
		\hline
		Half space ($qd\gg1$)
		& $\displaystyle {\cal K}(q)=\frac{1-\nu}{G} \frac{1}{|q|}$ 
		& $\displaystyle u(x)=-\frac{1-\nu}{\pi G}\int\ln|x-x'|p(x')dx'$ \\ 
		\hline
		\begin{tabular}[c]{@{}l@{}}Thin layer ($qd\ll1$),\\ compressible ($\nu\neq\frac{1}{2}$)\end{tabular} 
		& $\displaystyle {\cal K}(q)= \frac{d(1-2\nu)}{2G(1-\nu)}$ 
		& $\displaystyle u(x)= \frac{d(1-2\nu)}{2G(1-\nu)} p(x)$ \\ 
		\hline
		\begin{tabular}[c]{@{}l@{}}Thin layer ($qd\ll1$),\\ incompressible ($\nu=\frac{1}{2}$)\end{tabular} 
		& $\displaystyle {\cal K}(q)= \frac{d^3}{3G} q^2$ 
		& $\displaystyle u(x)= \frac{d^3}{3G} \frac{d^2p}{d x^2}$ \\
		\hline
	\end{tabular}
	\caption{
		Overview of asymptotes for the Fourier space Green's function ${\cal K}(q)$, given in \eqref{eq:kernel}. The two columns give the expansions of the Green's function and the corresponding pressure-deformation relations in real space.
	}
	\label{tbl:elasticity_asymptotes}

\end{table}

\subsection{Regimes} \label{s:regimes}

Since our goal is to identify and describe the various limits of the elastohydrodynamic problem, the relevant length scales need to be carefully compared. These length scales as defined in figure \ref{fig:schematic} are given by the contact size $a$, the elastic layer thickness $d$, the boundary layer width $\ell$, the typical surface displacement $u_0$, and the sought-after lubricant thickness $h_0$. The reader can already consult table~\ref{tbl:Scaling} for a complete overview of the various regimes.

\textbf{Small versus large deformations.} Two key limits of the problem come from the comparison between the deformation $u_0$ and the fluid film thickness $h_0$. When the sliding velocity of the cylinder is high, the cylinder experiences a strong upward force due to viscous stresses that separates it from the substrate. This gives the \emph{small deformation limit}, where $h_0\gg u_0$~\citep[]{Skotheim2005a, pandey2016}. By contrast, for relatively low sliding velocity the cylinder indents the substrate with only a thin layer of fluid in between the two, so that $h_0\ll u_0$. This is what will be referred to as the \emph{large deformation limit}~\citep[]{Bissett1989,Snoeijer2013}; remark that we still work in the context of linear elasticity, so that typical strains remain small.

\textbf{Short versus long wavelengths.} In the large deformation limit, a narrow lubrication boundary layer develops at the edge of the contact which regularizes the jump in pressure gradient caused by elasticity. Comparing the width of this boundary layer $\ell$ to the substrate thickness $d$, we can further distinguish two limits in the regime of large deformations. When $\ell\gg d$, the elastic problem is in the \emph{long wavelength limit}, in which the thin-layer approximation ($qd\ll1$) is valid over the full width of the contact. In the \emph{short wavelength limit} ($\ell\ll d$), the thin elastic layer approximation does not hold at the edge of the contact, and substrate locally responds like a half space ($qd\gg 1$) within the boundary layer. 

\textbf{Compressible versus incompressible elastic layers.} For each of the three above mentioned limits, one can choose a suitable response function given in table~\ref{tbl:elasticity_asymptotes}. 
For a comparatively thin layer, $d\ll \ell$, one needs to make an additional distinction between a compressible and an incompressible material. This distinction is made since the leading order term of the Taylor expansion of the kernel scales with $(1-2\nu)$, which vanishes in the incompressible case ($\nu=\frac{1}{2}$) making the next order term dominant. 

In the rest of the paper, we develop solutions of the elastohydrodynamic equations and in table~\ref{tbl:Scaling} we summarise these solutions through the scaling laws for the different regimes discussed above.

 \subsection{Non-dimensionalization}

While various of the cases reported in table~\ref{tbl:Scaling} are already available in the literature, we are not aware of any lubrication analysis of large deformations on thin elastic layers. Therefore, we will from now on consider lubricated contacts with thin elastic coatings, and we primarily focus on the case of compressible layers. We introduce dimensionless variables that are suitable for this specific case (in the long-wave limit of the elastic deformation), and identify the single dimensionless number that governs the solution. 

We consider the geometric relation (\ref{eq:Geometry_equation}) along with the deformation on thin compressible layers as given in table \ref{tbl:elasticity_asymptotes}. Combining these two equations we get,
\begin{equation} \label{eq:Deformation_Geometry_compressible_long_wave}
h(x)+c-\frac{x^2}{2R}=\frac{d\left(1-2\nu\right)}{2G (1-\nu)}p(x).
\end{equation}
The small deformation limit has been solved before in \cite{Skotheim2005a}, but here we consider the opposite limit. To non-dimensionalize the above equation for this limit we introduce the following scales,

\begin{align} \label{eq:Scaling_compressible_long_wave}
x=\bar x a, && 
h=\bar h u_0=\bar h \frac{a^2}{2R}, &&
p=\bar p\frac{3L}{4a}  ,
\end{align} 
where variables with an overbar are dimensionless. From this, one infers a natural characteristic horizontal scale, the contact size $a$, as
\begin{equation}\label{eq:scalingwidth}
a=\left[\frac{3}{4}\frac{LdR\left(1-2\nu\right)}{G \left(1-\nu\right)}\right]^{1/3},
\end{equation} 
so that \eqref{eq:Deformation_Geometry_compressible_long_wave} reduces to 

\begin{equation}
\bar{p}(\bar{x})=\bar{h}(\bar{x})+\bar c-\bar{x}^2.
\label{shape_nd}
\end{equation}
Here we further introduced $\bar c = c/u_0$, based on the natural vertical scale 
\begin{equation}
u_0 = \frac{a^2}{2R} = \left[\frac{3}{2^{7/2}}\frac{Ld\left(1-2\nu\right)}{R^{1/2}G \left(1-\nu\right)}\right]^{2/3}.
\end{equation}

Using the relation between pressure and deformation (\ref{shape_nd}), we find that Reynolds equation (\ref{eq:Reynolds_equation}) can be written as
\begin{equation} \label{eq:ODE_thin_compressible}
\pd{\bar{h}}{\bar{x}}-2\bar{x}=\lambda\frac{\bar{h}-\bar{h}^*}{\bar{h}^3},
\end{equation}
which contains the single dimensionless parameter of the problem,

\begin{equation} \label{eq:Lambda_thin_compressible}
\lambda=32\frac{V\eta}{L}\frac{R^2}{a^2}.
\end{equation} 
This parameter can be interpreted as the dimensionless velocity. The lubrication problem for a compressible elastic layer in the long-wave limit, is thus described by a first order ODE for the flowing thickness $\bar h(\bar x)$, given by \eqref{eq:ODE_thin_compressible}. Note, however, that there are two boundary conditions to be imposed, namely the pressure must vanish at $\bar x = \pm \infty$, which can be expressed in terms of $\bar h$ using (\ref{shape_nd}). The two boundary conditions for the first order ODE \eqref{eq:ODE_thin_compressible} can indeed be satisfied by tuning the value of $\bar h^*$. In addition, the coefficient $\bar c$ must be tuned such that 

\begin{equation}
\int_{-\infty}^{\infty}\bar{p}(\bar{x})\ d\bar{x}=\frac{4}{3},
\end{equation}
in order to impose the desired value of the load.

\subsection{Numerical methods} \label{s:numerical_methods}

To confirm the analytical results and to assess the crossover between the asymptotic regimes, we employ two numerical methods. The first method is numerically solving \eqref{eq:ODE_thin_compressible} in Wolfram Mathematica, along with the corresponding boundary conditions and constraints. We use a shooting method for a range of values for $\bar c$, finding the pair of $\bar h^*$ and $\lambda$ that allows to satisfy the boundary conditions, $\bar p(\bar x\to\pm\infty)=0$ and $\int^\infty_{-\infty} \bar p(\bar x)d\bar x = \tfrac{4}{3}$. Typical examples are given in fig.~\ref{fig:Unscaled_Profiles}. These accurately resolve the problem as long as the solution is self-consistent with the long-wave expansion. 

\begin{figure}
	\centering
	\includegraphics[width=\linewidth]{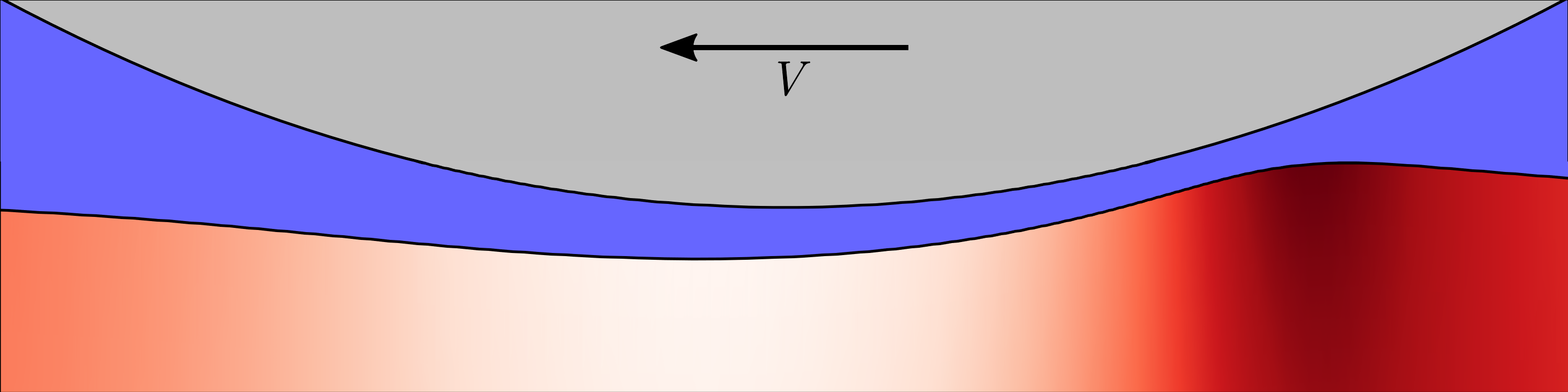}%

	\caption{
		Typical lubricated contact obtained in finite element simulations. The colourmap indicates the trace of the internal stresses in the thin, elastic layer. This result is obtained at an intermediate value of the sliding velocity, where the elastic deformation is comparable to the minimal lubrication thickness.}
	\label{fig:Fem2d}
\end{figure}

However, we will find that in the limit of very small $\lambda$, the obtained solution is no longer consistent with the long-wave expansion (of the elastic solid) that underlies \eqref{eq:ODE_thin_compressible}, we remark that the long-wave approximation for the liquid is satisfied for all cases considered. In this case, we turn to finite element simulations, where the two-dimensional plane strain problem is implemented using the library oomph-lib \citep{heil2006oomph}. A typical example of the numerical simulation is shown in figure \ref{fig:Fem2d}. The compressible elastic layer is simulated by a non-linear solid model using the Neo-Hookean constitutive law with a Poisson ratio of $\nu=0.3$. Since the strains in this problem always remain small, the simulations reduce to linear elasticity and the choice of the non-linear rheology does not influence the result. The domain has a width $W=10^5\ u_0$ ($|x| \leq W/2$) and height $d=30\ u_0$, and consists of refineable quad elements. The cylinder has a radius of $R=10^6\ u_0$, making for a contact size of $a\simeq1.4\times10^3\ u_0$. The fluid pressure is imposed on the top boundary of the elastic layer using surface elements that simulate the Reynolds equation \eqref{eq:Reynolds_equation}. The pressure in the fluid is constrained to zero at both sides of the domain and the nodes at the bottom of the elastic layer are fixed in both the vertical and horizontal direction. The nodes in both the left and right side of the layer are fixed only in the horizontal direction and are subject to a no-shear condition in the vertical direction.

\section{Thin compressible layer}\label{sec:thincomp}

In this section we present results for the lubricated contact with a thin, compressible, elastic substrate. We first illustrate the emergence of three distinct asymptotic regimes by presenting numerical results, both from the long-wave expansion of \eqref{eq:ODE_thin_compressible} and based on the finite element simulations. Then, we turn to a detailed asymptotic analysis of the two regimes that involve large deformations, respectively corresponding to long and short wavelengths.

\subsection{Numerical results}

\begin{figure}
	\centering
	
	\includegraphics[]{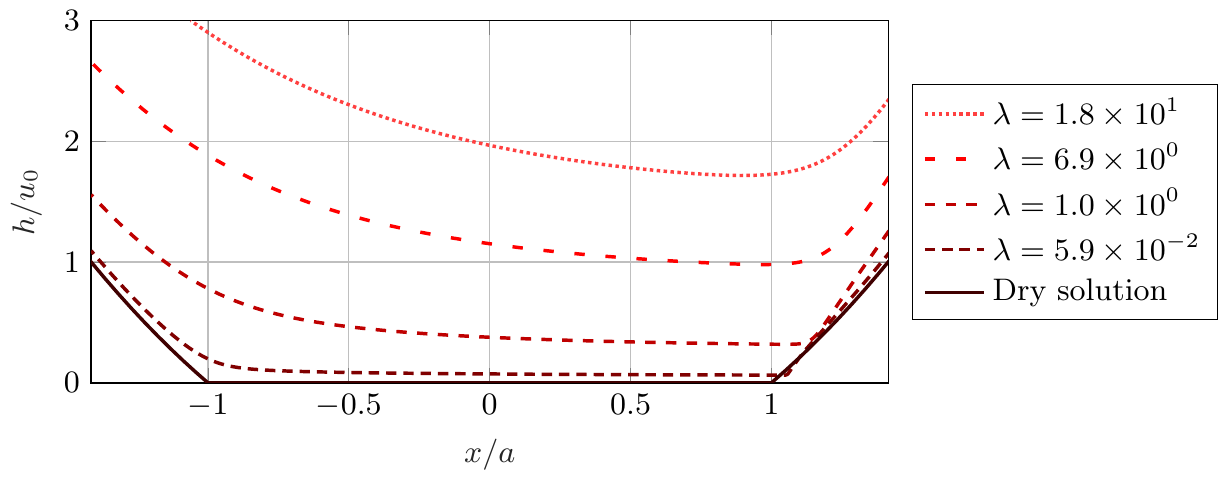}%
	
	\caption{
		Numerical film thickness profiles $\bar h = h/u_0$ as a function of $\bar x = x/a$, for the case of a compressible thin elastic layer. Lubricated contacts correspond to  $\lambda>0$, and are obtained from numerical integration of  \eqref{eq:ODE_thin_compressible}. The case $\lambda=0$ is a ``dry" contact, given by the analytical expression of \eqref{eq:Dry_compressible_long_wave_h}. 
	}
	\label{fig:Unscaled_Profiles}
\end{figure}

Figure \ref{fig:Unscaled_Profiles} reports a number of film thickness profiles $\bar h=h/u_0$ for lubricated contacts, for various values of the dimensionless sliding velocity $\lambda$ as defined by \eqref{eq:Lambda_thin_compressible}. The results presented in the figure were obtained from numerical integration of \eqref{eq:ODE_thin_compressible}, and also includes the ``dry" solution, corresponding to the case $\lambda=0$ where there is no motion. The profiles in fig.~\ref{fig:Unscaled_Profiles} clearly reveal a qualitative change in the shape of the contact, upon increasing the velocity. At small $\lambda$, the profiles exhibit a large elastic deformation where the cylinder pushes deep into the soft coating. The liquid profile closely resembles that of the ``dry" solution, except near the inlet ($\bar x=-1$) and the outlet ($\bar x=1$) regions where the effect of lubrication is prominent. At these two edges of the contact region, we encounter a narrow boundary layer of width $\ell$, as sketched in the inset of fig.~\ref{fig:schematic}. These boundary layers will be analysed in detail below, as they are essential for computing the entrained liquid thickness. At large $\lambda$, the liquid thickness becomes very large and the profile is smooth over the entire horizontal range. Gradually the gap approaches the shape of the rigid cylinder, since the elastic deformation of the coating becomes small in comparison to the lubricant thickness. 

\begin{figure}
	\centering
	
	\includegraphics[]{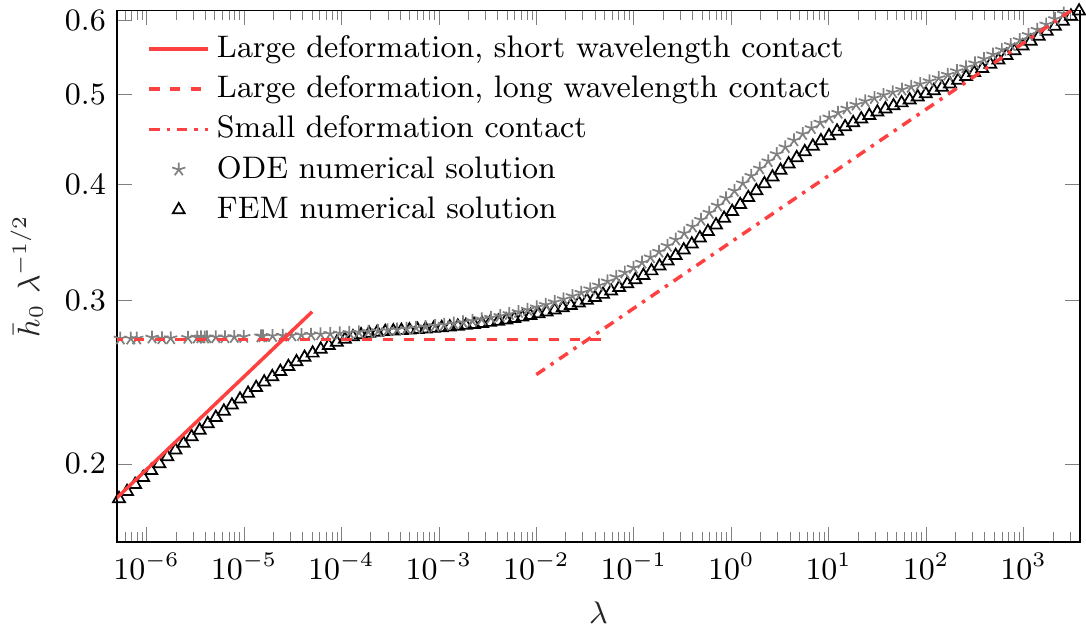}%
	
	\caption{
		Lubrication thickness for thin compressible coatings, combining numerical (Section \ref{s:numerical_methods}) and analytical (Table \ref{tbl:Scaling}, middle row) results. The graph reports the thickness of the fluid film at the center of the lubricated contact, $h_0$, as a function of the dimensionless sliding velocity $\lambda$. The dimensionless thickness $\bar h=h_0/u_0$ is compensated by the intermediate asymptotics $\bar h_0 \sim \lambda^{1/2}$ (dashed line). Distinct asymptotic limits are observed at small $\lambda$ (solid line, $\bar h_0 \sim \lambda^{3/5}$) and large $\lambda$ (dash-dotted line, $\bar h_0 \sim \lambda^{4/7}$). We remark that each of the indicated asymptotes are derived analytically (including the prefactors).
	}
	\label{fig:Thin_layer_compressible_combined}
\end{figure}

We now take a closer look at the deformations over the full range of $\lambda$ for the long-wave description and the finite element simulations -- both pertaining to thin compressible coatings. In figure \ref{fig:Thin_layer_compressible_combined} we therefore plot the fluid layer thickness at the center $\bar h_0=h_0/u_0$, as a function of the dimensionless velocity $\lambda$. The layer thickness is furthermore compensated by the scaling $\lambda^{1/2}$ that will be derived analytically below. From the figure we can see that for values of $\lambda\gtrsim10^{-4}$ the two different numerical methods agree with one another -- confirming the validity of the long-wave expansion at moderate and large values of $\lambda$. The long-wave data (represented as stars), exhibit two asymptotic regimes: $h_0 \sim \lambda^{1/2}$ (dashed line) and $h_0 \sim \lambda^{4/7}$ (dash-dotted line). However, it is also clear that at the smallest $\lambda$, the long-wave description breaks down, since the numerical solution of the true elastohydrodynamic problem (represented by triangles) gives yet another scaling law $h_0 \sim \lambda^{3/5}$ (solid line). We will see below that this limit emerges when the width of the boundary layer $\ell$ becomes comparable to the thickness of the coating $d$, so that the long-wave description is no longer justified. 

In summary, we thus find the emergence of 3 regimes at small, intermediate and large values of $\lambda$. The analysis for large $\lambda$ has already been performed by \citet{Skotheim2004,Skotheim2005a}. Hence, below we derive the intermediate asymptotics (using the long-wave description at small $\lambda$), and the ultimate asymptotics $\lambda \ll 1$ based on the short-wave description.

\subsection{Long wavelength elasticity: Dry solution}

Before turning to the dynamical case, we first provide the dry (static) solution of the long-wave approximation, given by \eqref{shape_nd}. Formally, this corresponds to $\lambda=0$, for which no flow takes place. This dry solution therefore corresponds to the situation where the pressure vanishes outside the contact region, $\bar{p}=0$ for $|\bar{x}|>1$, and the gap thickness vanishes within the contact, $\bar{h}=0$ for $|\bar x|<1$. The corresponding dry solution to \eqref{shape_nd} reads,

\begin{subequations}
	\begin{align} \label{eq:Dry_compressible_long_wave_p}
	\bar{p}&=\left(1-\bar{x}^2\right)\Theta\left(1-\bar{x}^2\right) \\
	\bar{h}&=\left(\bar{x}^2-1\right)\Theta\left(\bar{x}^2-1\right), \label{eq:Dry_compressible_long_wave_h}
	\end{align}
\end{subequations}
where $\Theta$ is the Heaviside step function. 

Expanding these near the edges of the contact, located at $\bar{x}=\mp 1$, we find

\begin{align} \label{eq:Matching_compressible_long_wave}
\bar{p}\simeq \pm 2(\bar{x}\pm 1) && \bar{h}\simeq \mp 2(\bar{x}\pm 1).
\end{align}
This linear behaviour of the gap can be observed in figure \ref{fig:Unscaled_Profiles}, at the edge of the dry solution. This asymptotics is of particular importance to the dynamical case at small speed, where $\lambda$ is small but nonzero. In that limit, lubrication effects quickly decay outside the narrow boundary layers near the edge of the contact, so that the static solution is approached. Specifically, the boundary layer solutions must therefore be matched to the linear asymptotics given by  (\ref{eq:Matching_compressible_long_wave}).

\subsection{Long wavelength elasticity: Lubricated solution} \label{s:Large_wl_comp_lubricated}

Once the cylinder starts to move, a thin film of liquid is entrained within the contact region and in particular, the left-right symmetry of the deformation is broken (fig.~\ref{fig:Unscaled_Profiles}). Liquid is squeezed into the contact at the inlet, at $\bar x=-1$, and is pushed out at the outlet, at $\bar x=1$.  To analyze the behavior of \eqref{eq:ODE_thin_compressible} at the two lubricated edges, we look for similarity solutions of the form 

\begin{align} \label{eq:Similarity_definition}
\bar{h}=\lambda^\alpha H(\xi) && \xi=(\bar{x}\pm 1)\lambda^{-\beta}. 
\end{align}Here the $+$ sign refers to the inlet and $-$ to the outlet. In terms of the similarity variables, the matching condition \eqref{eq:Matching_compressible_long_wave} becomes $H(\xi) \simeq \mp 2\xi$ for $\xi \rightarrow \mp \infty$. This $\lambda$-independent far-field condition requires that $\alpha=\beta$. The Reynolds lubrication equation  \eqref{eq:ODE_thin_compressible} written in terms of the above similarity variables, gives

\begin{equation}
\lambda^{\alpha-\beta}\pd{H}{\xi}\pm 2-2\lambda^\beta\xi=\lambda^{1-2\alpha}\frac{H-H^*}{H^3}.
\label{seq1}
\end{equation}
Anticipating that $\beta>0$, i.e. thin boundary layers for $\lambda\ll1$, the term $\sim \lambda^{\beta}$ is subdominant with respect to the constant. For the remaining equation to be independent of $\lambda$, we need $\alpha-\beta=1-2\alpha$. Combined with the matching condition of $\alpha=\beta$, we thus find the exponents $$\alpha=\beta=\frac{1}{2},$$ while \eqref{seq1} reduces  to 

\begin{equation} \label{eq:similarity_eqn}
\pd{H}{\xi}\pm 2=\frac{H-H^*}{H^3}.
\end{equation}
The solution to this equation must match to $H(\xi \rightarrow \mp \infty) = \mp 2\xi$, as discussed before. Furthermore, the gap should converge to a film of constant thickness inside the contact, leading to $\pd{H}{\xi}(\xi \rightarrow \pm \infty) = 0$.

\subsubsection{Inlet region}

At the inlet \eqref{eq:similarity_eqn} becomes
\begin{equation}\label{eq:sim1}
\pd{H}{\xi}=\frac{H-H^*-2 H^3}{H^3}.
\end{equation}
We first wish to find the constant thickness $H_0$ that is approached when entering the central zone of the contact (corresponding to $\xi\rightarrow\infty$). Interestingly, $H_0$ is not simply equal to $H^{\ast}$, but is to be determined from the equation $H_0-H^*-2 H_0^3=0$. For this film thickness to take on a positive real value, it is required that $H^* \geq \frac{1}{9}\sqrt{6}$. However, under this condition there are still two possible solution branches of $H_0$ for a given $H^*$.

For the selection of the relevant $H_0$, we now investigate whether the solution indeed approaches $H_0$ for $\xi\rightarrow\infty$. We therefore linearise \eqref{eq:sim1} using $H=H_0 + H_1(\xi)$, which gives

\begin{equation}
\pd{H_1}{\xi}=\frac{1-6 H_0^2}{H_0^3} H_1.
\end{equation}
This shows that the solution exponentially approaches a constant film thickness when $H_0 > \frac{1}{6}\sqrt{6}$. This condition selects the branch of acceptable $H_0$, which for a given value of $H^*$ indeed decays to a flat film below the contact. We need to separately consider the special case where $H_0 = \frac{1}{6}\sqrt{6}$ and $H^* = \frac{1}{9}\sqrt{6}$. In this case the similarity solution can even be found in closed form,

\begin{equation}\label{eq:solinlet}
\xi =  -\frac{1}{2}H + \frac{4}{9}H_0\ln\left (\frac{\frac{1}{2}H+H_0}{H+H_0}\right) + \frac{1}{36\left( H-H_0\right) }.
\end{equation} 
This solution exhibits an algebraic decay to the flat film of thickness $H_0$ as $\xi \rightarrow \infty$, and therefore should also be considered.

It will turn out that, indeed, the special case of $H_0 = \frac{1}{6}\sqrt{6}$ and $H^* = \frac{1}{9}\sqrt{6}$ provides \emph{the} relevant solution for the lubrication problem. This will be shown by analysing the ``central region", as done below in Sec.~\ref{subsec:central}. Hence, the correct inlet thickness reads

\begin{equation}
H_{\rm in} = \frac{1}{6}\sqrt{6} \approx 0.408,
\end{equation}
while the corresponding value for $H^*$ is given by 

\begin{equation}
H^* = \frac{1}{9}\sqrt{6} \approx 0.272.
\end{equation}

\subsubsection{Outlet region}

The similarity equation for the outlet differs from \eqref{eq:sim1} only in the sign of the last term on the right hand side,
\begin{equation}\label{eq:sim2}
\pd{H}{\xi}=\frac{H-H^*+2 H^3}{H^3}.
\end{equation}
Anticipating that $H^* = \frac{1}{9}\sqrt{6}$, we could find a lengthy closed form solution to (\ref{eq:sim2}) using Mathematica (not given here). 
Remarkably, the solution approaches a constant thickness $H_{\rm out}$ as $\xi \rightarrow -\infty$ that turns out to be different  from the inlet thickness $H_{\rm in}$. Indeed, solving $\pd{H}{\xi}=0$ for $H^*=\frac{1}{9}\sqrt{6}$, one finds 

\begin{equation}
H_{\rm out} = \frac{1}{6}\sqrt{6}\left[ \left(\sqrt{2}+1\right)^{\frac{1}{3}}-\left(\sqrt{2}-1\right)^{\frac{1}{3}}
\right]\approx 0.243,
\end{equation} 
which is significantly below $H_{\rm in}$. This means the liquid film cannot possibly be flat within the central region connecting the inlet and outlet, and we need to treat this region separately.

\subsubsection{Central region}\label{subsec:central}

As just demonstrated the constant film thickness at the inlet and outlet cannot be equal, these satisfy two different equations for the same value of $H^*$. To find the variation in film thickness across the central zone, another similarity solution has to be introduced, namely,

\begin{equation}
\bar h=\lambda^{1/2} {\cal H}(\bar{x}).
\end{equation}
Since this similarity solution does not scale the solution in horizontal direction, this equation is valid when $|\bar{x}|\lesssim1-\lambda^{1/2}$, i.e. over the full width of the contact \emph{excluding} the boundary layers. Substituting this into the lubrication equation \eqref{eq:ODE_thin_compressible} gives an algebraic equation for the film thickness,

\begin{equation}\label{eq:central}
{\cal H}- H^*+ 2\bar{x} {\cal H}^3=0, \quad \Longrightarrow \quad \bar x = \frac{H^*- {\cal H}}{2{\cal H}^3}.
\end{equation}
Since $\mathcal H$ describes the central zone up to the boundary layers, (\ref{eq:central}) needs to cover the entire domain $-1 < \bar x < 1$. One verifies that this is only the case when $H^* \leq \frac{1}{9}\sqrt{6}$. This condition is exactly opposite to the requirement derived from the inlet solution, which only gives physical solutions for $H^* \geq \frac{1}{9}\sqrt{6}$. Hence, the only possibility for the central zone to connect inlet and outlet is when $H^* = \frac{1}{9}\sqrt{6}$, as anticipated above.

\subsubsection{Summary and numerical validation}

\begin{figure}
	\centering
	
	\includegraphics[]{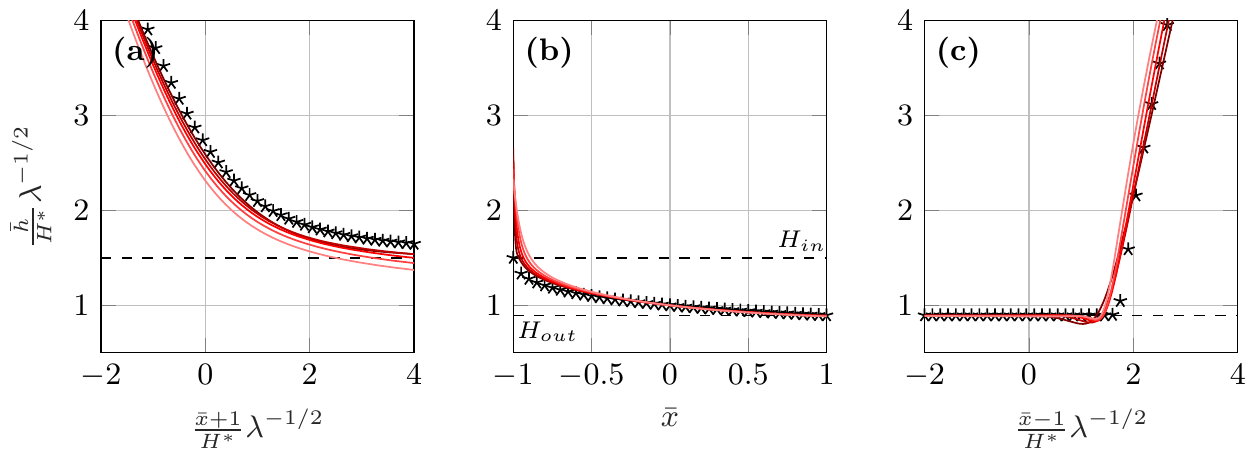}%
	\caption{
		Fluid film thickness profiles at the (a) inlet, (b) central region and (c) outlet of a lubricated contact on a compressible thin layer in the long wave limit. The solid lines represent rescaled FEM solutions for $4.99 \times 10^{-4} \leq \lambda \leq 2.26\times 10^{-2}$, while the symbols denote the similarity solution derived in section \ref{s:Large_wl_comp_lubricated}. The solution in (b) is plotted over the full range $|\bar{x}|<1$, but is not valid in the region near the edge of the contact.}
	\label{fig:Thin_layer_compressible_similarity2}
\end{figure}

Figure \ref{fig:Thin_layer_compressible_similarity2} summarizes the key results of this section. The panels (a-c) respectively report the similarity solutions for the narrow boundary layers at (a) the inlet, (c) the outlet and (b) the central region that connects the two. The symbols represent the analytical solutions derived above, and these are plotted along with scaled FEM numerical solutions (solid lines, for various $\lambda$). An excellent agreement is observed in all three regions. 

The similarity solution also provides the exact expression for the fluid film thickness $h_0$. For this we evaluate \eqref{eq:central} at $x=0$, which gives immediately that the (scaled) film thickness at the center of the contact is equal $H^* = \frac{1}{9}\sqrt{6}$. In dimensional form it gives

\begin{equation}\label{eq:h0longwave}
h_0= \frac{a^2}{2R} H^* \lambda^{1/2} = \left[\frac{2^{4}}{3^{7/2}}\sqrt{\frac{\eta^3 V^3}{G^2 L}}\frac{Rd(1-2\nu)}{1-\nu}\right]^{1/3},
\end{equation}
This result, including the theoretical prefactor, is used in fig.~\ref{fig:Thin_layer_compressible_combined} (dashed line), which shows that the numerical solutions of the long-wavelength approximation indeed converge towards the derived similarity solution. With respect to the full resolution of the elastohydrodynamic problem using finite elements, the solution (\ref{eq:h0longwave}) appears as an intermediate (long-wave) asymptotics. 

\subsection{Short wave elasticity: the ultimate asymptotics for $\lambda \ll 1$}\label{subsec:ultimate}

The numerical results obtained by finite element simulations, shown in fig.~\ref{fig:Thin_layer_compressible_combined}, revealed that at very small $\lambda$, another limit develops. The reason for this is that the boundary layer width in the long-wave theory was found to scale as $\ell \sim \lambda^{1/2}$, so that at sufficiently small $\lambda$ it becomes comparable to the thickness of the elastic layer -- at which point the long-wavelength description becomes inconsistent. Hence, the ultimate small $\lambda$ asymptotics involves the hierarchy of scales $\ell \ll d \ll a$, and requires the full kernel of \eqref{eq:kernel}. Below we first consider the corresponding static problem, to identify the appropriate horizontal, vertical and pressure scales. Subsequently, in the next subsection, we use the dry solution to match the boundary layers. 

\subsubsection{Dry solution}

\begin{figure}
\centering
\includegraphics[]{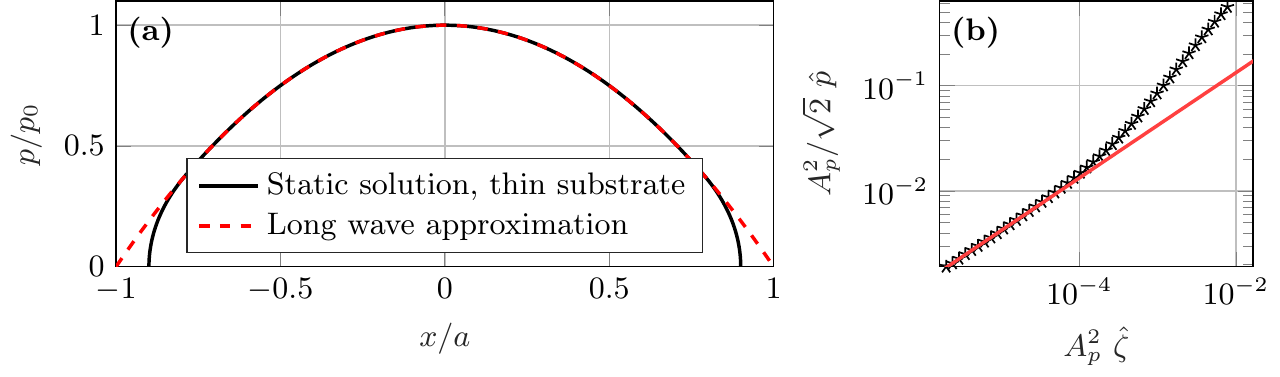}%
\caption{(a) Schematic representation of the pressure profile of a \emph{static} contact, on a thin compressible layer. The long-wave approximation (red dashed) breaks down near the edge of the contact, at a distance comparable to the thickness of the elastic layer (in this case $d \sim 0.1 a$, where $d$ is the layer thickness and $a$ the contact width). (b) Pressure profile at the edge of a dry, static contact on a thin compressible layer, resolved using finite element simulations (symbols). The data are fitted with \eqref{eq:pressure_halfspace_region_compressible} shown as the solid line, from which we obtain the numerical constant $A_p\approx1.33$ defined in (\ref{eq:Aconstant}). }
\label{fig:pressureprofiles}
\end{figure}

In fact, even the static indentation problem with $d \ll a$, already involves the full kernel \eqref{eq:kernel}. This problem  goes back to~\citet[]{meijers1968contact}, who considered the Hertz contact on layers of arbitrary thickness. On the scale of the contact width $a$, the static solution of the long-wavelength description, (\ref{eq:Dry_compressible_long_wave_p}, \ref{eq:Dry_compressible_long_wave_h}), is perfectly valid when $d\ll a$. Near the edges, however, the linear behaviour of the pressure gives way to a square-root dependence, as is illustrated in fig.~\ref{fig:pressureprofiles}(a). The reason for this is that near the edges, the elasticity is governed by the short-wavelength limit of the the elastic kernel \eqref{eq:kernel}. Introducing $\zeta=x\pm a$, indicating the unscaled distance to the contact edge, this gives the elastic relation

\begin{equation} \label{eq:Deformation_Geometry_compressible_short_wave_zeta}
h(\zeta)+c-\frac{\left( \zeta\mp a\right) ^2}{2R}=-\frac{1-\nu}{\pi G}\int_{-\infty}^{\infty}\ln|\zeta-\zeta'|p(\zeta')d\zeta',
\end{equation}
which combines \eqref{eq:Geometry_equation} and the deformation from table \ref{tbl:elasticity_asymptotes} in the limit $qd \gg 1$. It is well-known that this equation gives 

\begin{equation} \label{eq:pressure_halfspace_region_compressible}
p \simeq A \sqrt{\pm \zeta},
\end{equation}
at the contact edge \citep[]{Snoeijer2013}, where the scaling properties of $A$ remain to be determined. This behaviour is indeed manifestly different from the linear behaviour obtained from the long-wave expansion (\ref{eq:Matching_compressible_long_wave}), which in terms of a dimensional scaling law can be written as

 \begin{equation}\label{eq:bloep}
p \sim \pm \frac{L}{a^2} \zeta,
\end{equation}
with $a$ given by (\ref{eq:scalingwidth}). 
To estimate the value of the constant $A$, we now use a matching of the two descriptions, as is depicted schematically in fig.~\ref{fig:pressureprofiles}(a). This is done by equating (\ref{eq:pressure_halfspace_region_compressible}) to (\ref{eq:bloep}) at a distance $|\zeta| \sim d$. This matching gives an expression for the parameter $A$,

\begin{equation}\label{eq:Aconstant}
A = A_p \left[\frac{G^2 L}{\sqrt{d} R^2} \frac{(1-\nu)^2}{(1-2\nu)^2}\right]^{1/3},
\end{equation}
where $A_p$ is a (dimensionless) constant. The value of the constant is obtained from a numerical solution using the finite element method, of the static dry contact problem. The result is given in fig.~\ref{fig:pressureprofiles}(b), from which we deduce that $A_p\approx 1.33$.

Now that we know the behavior of the static solution near the contact edge, we can identify the proper scalings for $p$, $h$ and $\zeta$. This is important, since these will subsequently be used to describe the dynamic boundary layers. To establish the scalings, we first take two derivatives of (\ref{eq:Deformation_Geometry_compressible_short_wave_zeta}), which gives 

\begin{equation} \label{eq:Deformation_Geometry_compressible_short_wave_zeta_scaled}
\pdd{h(\zeta)}{\zeta}=\frac{1}{R}-\frac{1-\nu}{\pi G}\int_{-\infty}^{\infty}\pd{p(\zeta')}{\zeta'}\frac{1}{\zeta-\zeta'}d\zeta',
\end{equation}
where we also performed an integration by parts. Demanding that the three terms in this equation are of the same order, while also respecting the scaling of $A$ in (\ref{eq:pressure_halfspace_region_compressible},\ref{eq:Aconstant}), we find the appropriate scalings to be 

\begin{subequations} \label{eq:scaling_comp_edge}
	\begin{alignat}{2} 
	p  &= &\frac{A_p^2}{2}&\left[\frac{G L^2}{Rd} \frac{(1-\nu)^7}{(1-2\nu)^4}\right]^{1/3} \hat{p}, \\ 
	\zeta   &=&\frac{A_p^2}{2}&\left[\frac{L^2 R^2}{G^2 d} \frac{(1-\nu)^{10}}{(1-2\nu)^4}\right]^{1/3} \hat{\zeta}, \\ 
	h  &= &\frac{A_p^4}{4}&\left[\frac{L^4 R}{G^4 d^2} \frac{(1-\nu)^{20}}{(1-2\nu)^8}\right]^{1/3} \hat{h}.
	\end{alignat}
\end{subequations}
With this, the dimensionless equation describing the elasticity of the layer reads

\begin{equation}\label{eq:Compressible_shortwave2}
\pdd{\hat{h}}{\hat{\zeta}}=1-\frac{1}{\pi}\int_{-\infty}^{\infty}\pd{\hat{p}(\hat{\zeta}')}{\hat{\zeta}}\frac{1}{|\hat{\zeta}-\hat{\zeta}'|}d\hat{\zeta}',
\end{equation}
while the asymptotics of the static solution reduces to
\begin{equation}\label{eq:psqrt}
\hat{p} \simeq \sqrt{\pm 2 \hat{\zeta}}.
\end{equation}

\subsubsection{Lubricated solution}

\begin{figure}
	\centering

	\includegraphics[]{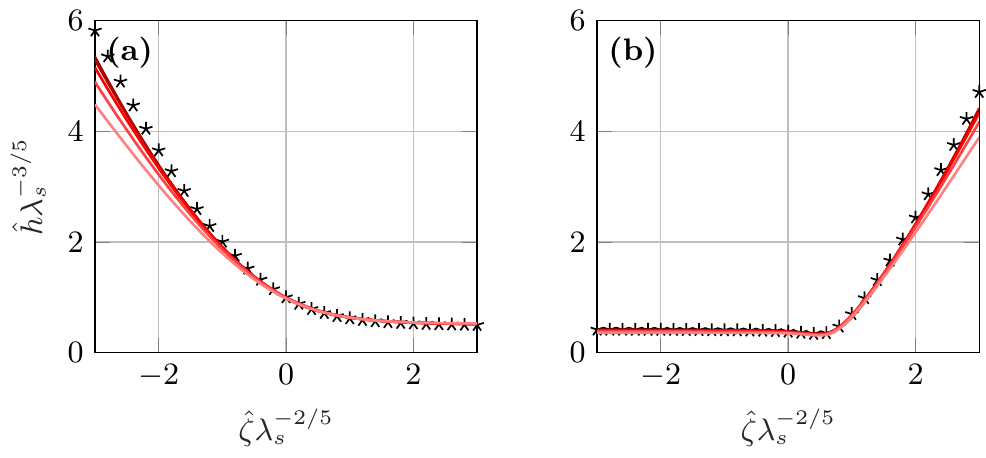}%
	\caption{Fluid film thickness profiles at the (a) inlet and (b) outlet of a lubricated contact on a compressible thin layer in the short wave limit (i.e. the ultimate regime for $\lambda\ll 1$). The solid line represents a rescaled FEM solution for $4.75 \times 10^{-7} \leq \lambda \leq 7.42\times 10^{-5}$, while the symbols show the similarity solutions by~\citet{Snoeijer2013}. These similarity solutions are shifted horizontally to align with the numerical solutions.}
	\label{fig:Thin_layer_compressible_similarity3}
\end{figure}

We now turn to the dynamical case, by expressing the Reynolds equation \eqref{eq:Reynolds_equation} in the new variables of \eqref{eq:scaling_comp_edge}. This gives

\begin{align} \label{eq:Compressible_shortwave1}
\pd{\hat{p}}{\hat{\zeta}}&=\lambda_s\frac{\hat{h}-\hat{h}^*}{\hat{h}^3},
\end{align}
which is of the same form as before. However, owing to the new scaling, the parameter that gives the dimensionless velocity is different, and reads

\begin{equation}
\lambda_s = \frac{96}{A_p^8} \left[\frac{\eta^3 V^3 d^4 R G^5}{L^8} \frac{(1-2\nu)^{16}}{(1-\nu)^{37}}\right]^{1/3}.
\end{equation}

The analysis of the boundary layer is now given by the system (\ref{eq:Compressible_shortwave2}, \ref{eq:Compressible_shortwave1}), with matching condition (\ref{eq:psqrt}). This reduced problem is strictly identical to that of the boundary layers on a half-space, as previously described by~\citet{Snoeijer2013}. Hence the solution is exactly the same, and gives a thickness $\hat h \sim \lambda_s^{3/5}$, a boundary layer width $\hat \zeta \sim \lambda_s^{2/5}$, while the pressure $\hat p \sim \lambda_s^{1/5}$. For detailed derivations we refer to~\citet{Snoeijer2013}. Note, however, that in dimensional variables the result is different. In particular, we find for the dimensional flowing thickness

\begin{align} \nonumber
h_0 &= 0.4467 \frac{A_p^4}{4} \left[\frac{L^4 R}{G^4 d^2} \frac{(1-\nu)^{20}}{(1-2\nu)^8}\right]^{1/3} \, \lambda_s^{3/5},\\
&=\frac{1.7271}{A_p^{4/5}} \left[ \frac{d^2 R^8 \eta^9 V^9 }{G^5 L^4} \frac{(1-2\nu)^8}{(1-\nu)^{11}}\right]^{1/15}, \label{eq:scaling_h_comp_sw}
\end{align}
where the numerical constant 0.4467 was determined in~\citet{Snoeijer2013} by solving $\hat h^*$ from the boundary layer problem, while we remind that $A_p\approx 1.33$.  This result is given as the solid line in fig.~\ref{fig:Thin_layer_compressible_combined}, and indeed accurately describes the ultimate $\lambda \ll 1$ asymptotics. 

The structure of the boundary layers is further illustrated in fig.~\ref{fig:Thin_layer_compressible_similarity3}. There, the similarity solutions from~\citet{Snoeijer2013} are reproduced (symbols), and compared to the scaled FEM simulations at very small $\lambda$. All curves indeed collapse, confirming the validity of the similarity analysis. Interestingly, for this case the  thickness of the lubrication layers leaving the inlet and entering the outlet directly align. The central zone is therefore perfectly flat -- this is to be contrasted with the result for the long-wave similarity solutions of fig.~\ref{fig:Thin_layer_compressible_similarity2}, for which the inlet and outlet were connected by a nontrivial central region.

\section{Discussion}\label{sec:discuss}

The aim of this work is to present all scaling laws governing the movement of a cylinder over an elastic substrate in the presence of a fluid. Some of these scaling laws were derived previously in the literature, but focussing on separate regimes -- importantly, the limiting cases of thin elastic layers with large deformations had not previously been considered. This important gap is filled in the present paper, thereby offering a complete and comprehensive overview.

Table \ref{tbl:Scaling} summarises both previous and our newly obtained results. All the different regimes of soft lubrication are organised in terms of the length scales $d$ (thickness of the elastic coating), $a$ (size of the contact), $u_0$ (scale of elastic deformation) and $\ell$ (boundary layer width near the contact edges). In terms of experimental parameters, the table should be read as follows. For fixed material properties, magnitude of the applied load $L$ determines the typical size of the contact $a$. Hence, upon specifying the load one selects one of the rows of table \ref{tbl:Scaling}, comparing the contact size to the elastic layer thickness. Then, at very small velocity $V$ one starts in the left column, where the flowing thickness $h_0 \ll u_0$, referring to the case of relatively large elastic deformation. Subsequently, upon increasing the velocity, the boundary layer $\ell$ and the lubrication film $h_0$ increases, and one progressively moves to the second and third columns, ultimately reaching the small deformation limit ($h_0 \gg u_0$).

\begin{table}
	\centering
	
	\begin{tabular}{llll} \hline
		& \multicolumn{2}{c}{\textbf{Large deformation ($\textstyle \bm{u_0\gg h_0}$)}} 
		& \multicolumn{1}{c}{\textbf{Small deformation}}
		\\ 
		& \multicolumn{1}{c}{$\textstyle \bm{\ell\ll d}$}               
		& \multicolumn{1}{c}{$\textstyle \bm{\ell\gg d}$}
		& \multicolumn{1}{c}{\textbf{($\textstyle \bm{u_0\ll h_0}$)}}
		\\ \hline
		$\bm{d\gg a}$
		& \makecell{$\displaystyle 0.603 \left[\frac{\eta^3 V^3}{G^2 L } R^3 (1{-}\nu)^2\right]^\frac{1}{5}$ \\ \citep{Snoeijer2013}}
		& \makecell{\textit{Not physically possible} \\ \textit{($\ell \lesssim a$)}}
		& \makecell{$\displaystyle \left[\frac{3\pi}{8}\frac{\eta^2 V^2}{G L } R^2(1{-}\nu)\right]^\frac{1}{3}$  \\ \citep{pandey2016}}      
		\\ \hline	
		$\begin{aligned} 
		&\textstyle \bm{d\ll a} \\ 
		&\textstyle \bm{\nu\neq\frac{1}{2}}
		\end{aligned}$
		& $\displaystyle {\sim}\left[ \frac{\eta^9 V^9 }{G^5 L^4} R^8 d^2 \frac{(1{-}2\nu)^8}{(1{-}\nu)^{11}}\right]^\frac{1}{15}$    
		& $\displaystyle \left[\frac{2^{8}}{3^{7}} \frac{\eta^3 V^3}{G^2 L} R^2 d^2 \frac{(1{-}2\nu)^2}{(1{-}\nu)^2} \right]^\frac{1}{6} $     
		& \makecell{$\displaystyle \left[\frac{6\pi}{8\sqrt{2}}\frac{\eta^2 V^2}{GL}R^{\frac{3}{2}}d\frac{1{-}2\nu}{1{-}\nu}\right]^\frac{2}{7}$   \\ \citep{Skotheim2005a}}         
		\\ \hline
		$\begin{aligned}
		&\textstyle \bm{d\ll a} \\ 
		&\textstyle \bm{\nu=\frac{1}{2}}
		\end{aligned}$
		& $\displaystyle {\sim}\left[ \frac{\eta^{15} V^{15}}{G^7 L^8} R^{12} d^6 \right]^\frac{1}{25}$          
		& \makecell{\textit{no long-wave} \\ \textit{asymptotics}  }
		& \makecell{$\displaystyle \left[\frac{15\pi}{8\sqrt{2}}\frac{\eta^2 V^2 }{GL} R^{\frac{1}{2}} d^3\right]^\frac{2}{9}$  \\ \citep{Skotheim2004}}         
		\\ \hline
	\end{tabular}
	\caption{Scaling laws for the film thickness at the center of a cylinder, denoted $h_0$, in the various regimes of soft lubrication. The limits are expressed in terms of length scales, as defined in fig.~\ref{fig:schematic}. Previously available results have been indicated by the corresponding reference. Results without a reference have been derived in the present manuscript.}
	\label{tbl:Scaling}
\end{table}

There are a few important features in this table. At very small velocities (left column), the scaling $h_0 \sim V^{3/5}$ is the same for all cases. In this case, the width of the boundary layer is the smallest scale in the system and thereby these exhibit a universal structure. Differences arise only from matching these boundary layers to different (static) outer solutions. On the contrary, at very large velocity (right column) the scaling laws always involve the combination $(\eta V)^2/L$. So, when working at a fixed separation distance instead of at fixed load, one finds the well known scaling law for the lift force $L \sim (\eta V)^2$~\citep{Sekimoto1993, Skotheim2004,Greenwood2020}. A key result of our analysis is the emergence of an intermediate asymptotics, $h_0\sim V^{1/2}$ for the thin, compressible layer -- closed form analytical solutions for this case are provided. 

\begin{figure}
	\centering
	
	\includegraphics[]{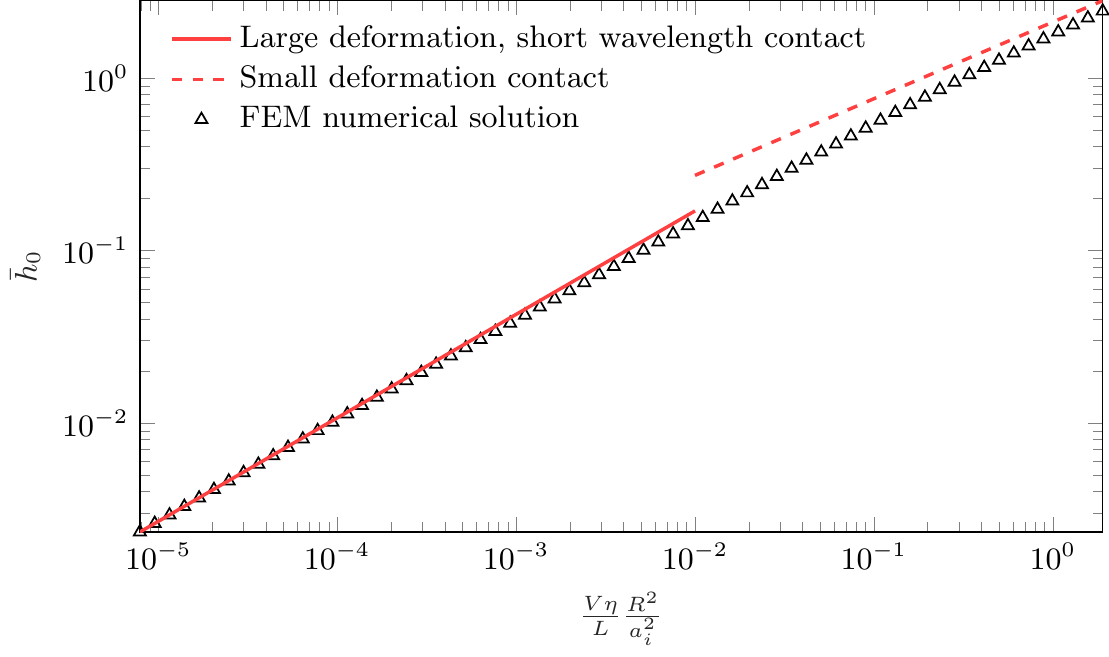}%
	
	\caption{
		Lubrication thickness for thin incompressible coatings, combining numerical (Section \ref{s:numerical_methods}) and analytical (Table \ref{tbl:Scaling}, middle row) results. The graph reports the thickness of the fluid film at the center of the lubricated contact, $h_0$, rescaled according to \eqref{eq:Scaling_incompressible_long_wave} as a function of a dimensionless sliding velocity. The dimensionless sliding velocity uses $a_i$ as defined in \eqref{eq:a_i}. Distinct asymptotic limits are observed at low velocity (solid line, $\bar h_0 \sim \lambda^{3/5}$) and high velocity (dashed line, $\bar h_0 \sim \lambda^{4/9}$). The prefactor of the high velocity asymptote is derived analytically, while for the low velocity asymptote the data was fitted by (\ref{eq:broeps}), with $A_{i,p}\simeq1.85$.
	}
	\label{fig:Thin_layer_incompressible_combined}
\end{figure}

Another new result of our study is the scaling law for a thin \emph{incompressible} layer subjected to large indentation. Leaving the details in appendix \ref{sss:dry_incompressible}, here we discuss the key features of this limit. In the limit where the boundary layer represents the smallest scale, $\ell \ll d$, the only change compared to the compressible case is in the static ``dry" solution, leading to different exponents of geometric and material parameters. However, the scaling $h_0 \sim V^{3/5}$ is robust. Next we consider the intermediate range, $d\ll \ell \ll a$, where thesmall boundary layers are in between the coating thickness and the contact size. Interestingly, this case does not admit a consistent long-wave description, and therefore a proper intermediate asymptotic regime appears to be missing. The resulting long-wavelength ODE can be derived -- unlike the compressible case, however, it does not admit solutions at small $\lambda$ that connect the inlet and the outlet region. For completeness, we show in fig.~\ref{fig:Thin_layer_incompressible_combined} the numerical relation between the central thickness $h_0$ and the velocity $V$ obtained by finite element simulations. It very nicely confirms the scalings of the bottom row in table \ref{tbl:Scaling}, corresponding to the small and large speed asymptotics for thin incompressible layers.

In conclusion, our study identifies key geometric parameters in the soft lubrication problem of compressible and incompressible elastic solids. 
From an experimental perspective, our results point out the importance of: (i) the ratio of elastic coating thickness to contact size, (ii) the Poisson ratio of the layer (compressible versus incompressible), (iii) the ratio of elastic deformation over the lubricant thickness. When interpreting experiments, one must verify the relevant regime, with Table \ref{tbl:Scaling} serving as a guide. Our unified framework also connects the two most studied limits of the problem: the lubricated Hertzian contact regime, relevant for many industrial applications, and the large separation (small deformation) regime, important in microfluidic and biological contexts. We anticipate that the results presented here to have important implications beyond the realm of soft lubrication, e.g. in understanding the rheology of suspension flow in channels~\cite[]{Rosti19}, size dependent migration of particles~\cite[]{Rallabandi18}, and direct measurement of interfacial rheology in soft matter systems~\cite[]{Garcia16}.\\

\noindent {\bf Acknowledgements:~} We thank Thomas Salez for numerous discussions on various limits of soft lubrications. We acknowledge financial support from ERC (the European Research Council) Consolidator Grant No. 616918, and from NWO through VICI Grant No. 680-47-632.\\

\noindent {\bf Declaration of Interests}. The authors report no conflict of interest.

\appendix
\section{Thin incompressible layer, small velocity} \label{sss:dry_incompressible}

The derivation of the dry contact between a cylinder on and an incompressible thin layer is similar to that of a compressible layer. The main difference lies in the elastic Green's function of table \ref{tbl:elasticity_asymptotes}, which differs in the long-wave limit. Combined with \eqref{eq:Geometry_equation}, the elastic response gives

\begin{equation} \label{eq:elasticity_thin_incompressible}
h(x) - \frac{x^2}{2R} + c= \frac{d^3}{3G}\pdd{p}{x}.
\end{equation}
In principle, this relation can be combined with the Reynolds equation \eqref{eq:Reynolds_equation} to yield an ODE for $h(x)$. It turns out, however, that this long-wave expansion does not admit solutions at small $\lambda$ that connect the inlet and the outlet region. Therefore, we only focus on the ultimate asymptotics at $\lambda \ll 1$, where the boundary layers width $\ell \ll d$ and are described by the short-wave kernel. The subsequent analysis now runs parallel to that of Section~\ref{subsec:ultimate}. The boundary layers exhibit a universal structure, but the corresponding scales need to be identified from matching to the static solution (now, on thin incompressible layers). In particular, we know the asymptotics to be of the form

\begin{equation}\label{eq:babbab}
p \simeq A_i \sqrt{\pm \zeta},
\end{equation}
where $\zeta=x\pm \chi a_i$ is the distance from the edge, and we need to identify $A_i$ of the static solution.

For this, we proceed as in Section~\ref{subsec:ultimate}. We first consider the static long-wave solution, from \eqref{eq:elasticity_thin_incompressible}, which gives the relevant horizontal length scale $a_i$,

\begin{equation} \label{eq:a_i}
a_i=\left[\frac{5}{6\sqrt{3}}\frac{Ld^3R}{G}\right]^{1/5}.
\end{equation}
We introduce the scalings,

\begin{align} \label{eq:Scaling_incompressible_long_wave}
p=\frac{5\sqrt{3}L}{16a_i} \bar{p}, && 
x=a_i \bar{x}, && 
h=\frac{a_i^2}{2R} \bar{h}.
\end{align}
As before the dry problem can be split into two distinct regions, which gives the conditions $\bar{h} = 0$ where $|\bar{x}|<\chi$ and $\bar{p} = 0$ where $|\bar{x}|>\chi$. Note that the size of the contact is not given by $a_i$, but by $\chi a_i$. This is because incompressibility causes the contact to be larger than the length over which the cylinder is below the surface of the undeformed layer. Incompressibility dictates that the volume of the layer is conserved,thus $\int_{-\infty}^\infty \left[\bar{h}-(\bar{x}^2-1)\right] d\bar{x} = 0$, which gives that the size of the contact $\chi=\sqrt{3}$. Using this result the dry solution is found to be,
\begin{subequations}
	\begin{alignat}{2}
	\label{eq:incomp_dry_h} \bar{h}&=\left(\bar{x}^2-1\right)&&\Theta\left(\bar{x}^2-\chi^2\right) \\
	\label{eq:incomp_dry_p} \bar{p}&=\tfrac{1}{9}\left(9-6\bar{x}^2+\bar{x}^4\right)&&\Theta\left(\chi^2-\bar{x}^2\right).
	\end{alignat}
\end{subequations}
Interestingly, while the pressure is a continuous function, the film thickness is a discontinuous function. Equation \eqref{eq:elasticity_thin_incompressible} predicts that the surface of the layer follows the cylinder when $|\bar{x}|<\chi$, and then suddenly becomes flat. 

We now need to compare the ``true" asymptotics of the statics solution, given by (\ref{eq:babbab}), to that of the long-wave approximation, which near the edge of the contact ($\bar{x}=\mp\chi$) is given by

\begin{equation} \label{eq:Dry_incompressible_expansion}
\bar{p}\simeq \pm\tfrac{4}{3}\left(\bar{x}\pm\chi\right)^2,
\end{equation}
which in dimensional form gives the scaling,

\begin{equation}
p \sim \pm \frac{L}{a_i^3} \zeta^2.
\end{equation}
Equating this expression to (\ref{eq:babbab}) at a distance $|\zeta| \sim d$, we find 

\begin{equation}
A_i = A_{i,p} \left[\frac{G^3 L^2}{d^{3/2} R^3} \right]^{1/5},
\end{equation}
where $A_{i,p}$ is a numerical constant that remains to be determined. To recover the ``universal" boundary layer equations, we scale the problem as,

\begin{subequations} \label{eq:scaling_incomp_edge}
	\begin{alignat}{2}
	p &=& \frac{A_{i,p}^2}{2} &\left[\frac{G L^4}{d^3 R}\right]^{1/5} \hat{p}, \\ 
	\zeta &=& \frac{A_{i,p}^2}{2} &\left[\frac{L^4 R^4}{G^4 d^3}\right]^{1/5} \hat{\zeta}, \\ 
	h &=& \frac{A_{i,p}^4}{4} &\left[\frac{L^8 R^3}{G^8 d^6} \right]^{1/5} \hat{h}.
	\end{alignat}
\end{subequations}
With this, the elastic deformation and the Reynolds equations reduce to \eqref{eq:Compressible_shortwave1} and \eqref{eq:Compressible_shortwave2}, but with a different dimensionless velocity,

\begin{equation}
\lambda_{s,i} = \frac{96}{A_{i,p}^8} \left[\frac{\eta^5 V^5 G^{11} d^{12}}{L^{16} R} \right]^{1/5}
\end{equation}
Therefore, the resulting central  thickness of the fluid film is therefore given by,
\begin{align} \nonumber
h_0&=0.4467\ \lambda_{s,i}^{3/5} \ \frac{A_{i,p}^4}{4} \left[\frac{L^8 R^3}{G^8 d^6} \right]^{1/5}\\
&=\frac{1.7271}{A_{i,p}^{4/5}} \left[ \frac{\eta^{15} V^{15} d^6 R^{12} }{G^7 L^8} \right]^{1/25}.\label{eq:broeps}
\end{align}

\bibliography{Soft_Lubrication}

\begin{thebibliography}{39}
\expandafter\ifx\csname natexlab\endcsname\relax\def\natexlab#1{#1}\fi
\def\au#1{#1} \def\ed#1{#1} \def\yr#1{#1}\def\at#1{#1}\def\jt#1{\textit{#1}}
  \def\bt#1{#1}\def\bvol#1{\textbf{#1}} \def\vol#1{#1} \def\pg#1{#1}
  \def\publ#1{#1}\def\arxiv#1{#1}\def\org#1{#1}\def\st#1{\textit{#1}}

\bibitem[Bissett(1989)]{Bissett1989}
{\sc \au{Bissett, E.~J.}} \yr{1989}  \at{The line contact problem of
  elastohydrodynamic lubrication. i. asymptotic structure for low speeds}.
  \jt{Proc. R. Soc. A}  \bvol{424}~(1867),  \pg{393--407}.

\bibitem[Bretherton(1961)]{Bretherton1961}
{\sc \au{Bretherton, F.~P.}} \yr{1961}  \at{The motion of long bubbles in
  tubes}.  \jt{J. Fluid Mech.}  \bvol{10},  \pg{166--188}.

\bibitem[Byun(2013)]{Byun2013a}
{\sc \au{Byun, S. et~al.}} \yr{2013}  \at{{Characterizing deformability and
  surface friction of cancer cells.}}  \jt{Proc. Natl. Acad. Sci. U.S.A.}
  \bvol{110}~(19),  \pg{7580--5}.

\bibitem[Cann {\em et~al.\/}(1996)Cann, Spikes \& Hutchinson]{CSH1996}
{\sc \au{Cann, P.~M.}, \au{Spikes, H.~A.} \& \au{Hutchinson, J.}} \yr{1996}
  \at{The development of a spacer layer imaging method (slim) for mapping
  elastohydrodynamic contacts}.  \jt{Tribol. T.}  \bvol{39}~(4),
  \pg{915--921}.

\bibitem[Davies {\em et~al.\/}(2018)Davies, D{\'e}barre, El~Amri, Verdier,
  Richter \& Bureau]{Davies18}
{\sc \au{Davies, H.~S.}, \au{D{\'e}barre, D.}, \au{El~Amri, N.}, \au{Verdier,
  C.}, \au{Richter, R.~P.} \& \au{Bureau, L.}} \yr{2018}
  \at{Elastohydrodynamic lift at a soft wall}.  \jt{Phys. Rev. Lett.}
  \bvol{120},  \pg{198001}.

\bibitem[Dowson(1998)]{dowson}
{\sc \au{Dowson, D.}} \yr{1998} {\em History of Tribology\/}.  \publ{Wiley, 2nd
  edition}.

\bibitem[Dowson \& Higginson(1959)]{DH1959}
{\sc \au{Dowson, D.} \& \au{Higginson, G.~R.}} \yr{1959}  \at{A numerical
  solution to the elasto-hydrodynamic problem}.  \jt{J, Mech. Eng. Sci.}
  \bvol{1}~(1),  \pg{6--15}.

\bibitem[Dowson \& Higginson(1966)]{DH1966}
{\sc \au{Dowson, D.} \& \au{Higginson, G.~R.}} \yr{1966} {\em
  Elasto-hydrodynamic Lubrication\/}.  \publ{Oxford: Pergamon Press}.

\bibitem[Fitz-Gerald(1969)]{Fitz1969}
{\sc \au{Fitz-Gerald, J.~M.}} \yr{1969}  \at{Mechanics of red-cell motion
  through very narrow capillaries}.  \jt{Proc. R. Soc. B}  \bvol{174}~(1035),
  \pg{193--227}.

\bibitem[Garcia {\em et~al.\/}(2016)Garcia, Barraud, Picard, Giraud, Charlaix
  \& Cross]{Garcia16}
{\sc \au{Garcia, L.}, \au{Barraud, C.}, \au{Picard, C.}, \au{Giraud, J.},
  \au{Charlaix, E.} \& \au{Cross, B.}} \yr{2016}  \at{A micro-nano-rheometer
  for the mechanics of soft matter at interfaces}.  \jt{Review of Scientific
  Instruments}  \bvol{87}~(11),  \pg{113906}.

\bibitem[Geislinger \& Franke(2014)]{GEISLINGER2014161}
{\sc \au{Geislinger, T.~M.} \& \au{Franke, T.}} \yr{2014}  \at{Hydrodynamic
  lift of vesicles and red blood cells in flow—from f{\aa}hr{\ae}us \&
  lindqvist to microfluidic cell sorting}.  \jt{Adv. Colloid Interface Sci.}
  \bvol{208},  \pg{161--176}.

\bibitem[Gohar \& Cameron(1963)]{CG1963}
{\sc \au{Gohar, R.} \& \au{Cameron, A.}} \yr{1963}  \at{Optical measurement of
  oil film thickness under elasto-hydrodynamic lubrication}.  \jt{Nature}
  \bvol{200}~(4905),  \pg{458--459}.

\bibitem[Greenwood(2020)]{Greenwood2020}
{\sc \au{Greenwood, J.~A.}} \yr{2020}  \at{Elastohydrodynamic lubrication}.
  \jt{Lubricants}  \bvol{8}~(5),  \pg{51}.

\bibitem[Hamrock \& Dowson(1977)]{HD1977}
{\sc \au{Hamrock, B.~J.} \& \au{Dowson, D.}} \yr{1977}  \at{Isothermal
  elastohydrodynamic lubrication of point contacts: part iii—fully flooded
  results}.  \jt{J. Lubr. Technol.} .

\bibitem[Hamrock {\em et~al.\/}(2004)Hamrock, Schmid \&
  Jacobson]{hamrock2004fundamentals}
{\sc \au{Hamrock, B.~J.}, \au{Schmid, S.~R.} \& \au{Jacobson, B.~O.}} \yr{2004}
  {\em Fundamentals of Fluid Film Lubrication\/}. {\em Mechanical
  engineering\/} .  \publ{CRC Press}.

\bibitem[Hannah(1951)]{hannah1951contact}
{\sc \au{Hannah, M.}} \yr{1951}  \at{Contact stress and deformation in a thin
  elastic layer}.  \jt{Q. J. Mech. Appl. Math.}  \bvol{4}~(1),  \pg{94--105}.

\bibitem[Heil \& Hazel(2006)]{heil2006oomph}
{\sc \au{Heil, M.} \& \au{Hazel, A.~L.}} \yr{2006}  \at{oomph-lib--an
  object-oriented multi-physics finite-element library}.  \bt{In {\em
  Fluid-structure interaction\/}},  \pg{pp. 19--49}.  \publ{Springer}.

\bibitem[Hou {\em et~al.\/}(1992)Hou, Mow, Lai \& Holmes]{Hou1992}
{\sc \au{Hou, J.~S.}, \au{Mow, V.~C.}, \au{Lai, W.~M.} \& \au{Holmes, M.~H.}}
  \yr{1992}  \at{An analysis of the squeeze-film lubrication mechanism for
  articular cartilage}.  \jt{J. Biomech.}  \bvol{25}~(3),  \pg{247 -- 259}.

\bibitem[Leroy {\em et~al.\/}(2012)Leroy, Steinberger, Cottin-Bizonne,
  Restagno, L\'eger \& Charlaix]{leroy2012}
{\sc \au{Leroy, S.}, \au{Steinberger, A.}, \au{Cottin-Bizonne, C.},
  \au{Restagno, F.}, \au{L\'eger, L.} \& \au{Charlaix, \'E.}} \yr{2012}
  \at{Hydrodynamic interaction between a spherical particle and an elastic
  surface: A gentle probe for soft thin films}.  \jt{Phys. Rev. Lett.}
  \bvol{108},  \pg{264501}.

\bibitem[Meeker {\em et~al.\/}(2004)Meeker, Bonnecaze \& Cloitre]{Meeker2004}
{\sc \au{Meeker, S.~P.}, \au{Bonnecaze, R.~T.} \& \au{Cloitre, M.}} \yr{2004}
  \at{Slip and flow in soft particle pastes}.  \jt{Phys. Rev. Lett.}
  \bvol{92},  \pg{198302}.

\bibitem[Meijers(1968)]{meijers1968contact}
{\sc \au{Meijers, P.}} \yr{1968}  \at{The contact problem of a rigid cylinder
  on an elastic layer}.  \jt{Appl. Sci. Res.}  \bvol{18}~(1),  \pg{353--383}.

\bibitem[Pandey {\em et~al.\/}(2016)Pandey, Karpitschka, Venner \&
  Snoeijer]{pandey2016}
{\sc \au{Pandey, A.}, \au{Karpitschka, S.}, \au{Venner, C.H.} \& \au{Snoeijer,
  J.~H.}} \yr{2016}  \at{Lubrication of soft viscoelastic solids}.  \jt{J.
  Fluid Mech.}  \bvol{799},  \pg{433--447}.

\bibitem[Rallabandi {\em et~al.\/}(2018)Rallabandi, Oppenheimer, Zion \&
  Stone]{Rallabandi18}
{\sc \au{Rallabandi, B.}, \au{Oppenheimer, N.}, \au{Zion, M. Y.~B.} \&
  \au{Stone, H.~A.}} \yr{2018}  \at{Membrane-induced hydroelastic migration of
  a particle surfing its own wave}.  \jt{Nature Physics}  \bvol{14}~(12),
  \pg{1211--1215}.

\bibitem[Rallabandi {\em et~al.\/}(2017)Rallabandi, Saintyves, Jules, Salez,
  Sch\"onecker, Mahadevan \& Stone]{Rallabandi17}
{\sc \au{Rallabandi, B.}, \au{Saintyves, B.}, \au{Jules, T.}, \au{Salez, T.},
  \au{Sch\"onecker, C.}, \au{Mahadevan, L.} \& \au{Stone, H.~A.}} \yr{2017}
  \at{Rotation of an immersed cylinder sliding near a thin elastic coating}.
  \jt{Phys. Rev. Fluids}  \bvol{2},  \pg{074102}.

\bibitem[Reynolds(1886)]{Reynolds1886}
{\sc \au{Reynolds, O.}} \yr{1886}  \at{{On the theory of lubrication and its
  application to Mr. Beauchamp Tower's experiments, including an experimental
  determination of the viscosity of olive oil}}.  \jt{Philos. Trans. R. Soc.
  London}  \bvol{177},  \pg{157}.

\bibitem[Rosti {\em et~al.\/}(2019)Rosti, Ardekani \& Brandt]{Rosti19}
{\sc \au{Rosti, M.~E.}, \au{Ardekani, M.~N.} \& \au{Brandt, L.}} \yr{2019}
  \at{Effect of elastic walls on suspension flow}.  \jt{Phys. Rev. Fluids}
  \bvol{4},  \pg{062301}.

\bibitem[Saintyves {\em et~al.\/}(2016)Saintyves, Jules, Salez \&
  Mahadevan]{Saintyves5847}
{\sc \au{Saintyves, B.}, \au{Jules, T.}, \au{Salez, T.} \& \au{Mahadevan, L.}}
  \yr{2016}  \at{Self-sustained lift and low friction via soft lubrication}.
  \jt{Proc. Natl. Acad. Sci. U.S.A.}  \bvol{113}~(21),  \pg{5847--5849}.

\bibitem[Salez \& Mahadevan(2015)]{Salez2015}
{\sc \au{Salez, T.} \& \au{Mahadevan, L.}} \yr{2015}  \at{Elastohydrodynamics
  of a sliding, spinning and sedimenting cylinder near a soft wall}.  \jt{J.
  Fluid Mech.}  \bvol{779},  \pg{181--196}.

\bibitem[Secomb {\em et~al.\/}(1986)Secomb, Skalak, {\"O}zkaya \&
  Gross]{Secomb1986}
{\sc \au{Secomb, T.~W.}, \au{Skalak, R.}, \au{{\"O}zkaya, N.} \& \au{Gross,
  J.~F.}} \yr{1986}  \at{Flow of axisymmetric red blood cells in narrow
  capillaries}.  \jt{J. Fluid Mech.}  \bvol{163},  \pg{405--423}.

\bibitem[Sekimoto \& Leibler(1993)]{Sekimoto1993}
{\sc \au{Sekimoto, K.} \& \au{Leibler, L.}} \yr{1993}  \at{{A Mechanism for
  Shear Thickening of Polymer-Bearing Surfaces : Elasto-Hydrodynamic Coupling
  .}}  \jt{Europhys. Lett.}  \bvol{23}~(2),  \pg{113--117}.

\bibitem[Skotheim \& Mahadevan(2004)]{Skotheim2004}
{\sc \au{Skotheim, J.~M.} \& \au{Mahadevan, L.}} \yr{2004}  \at{{Soft
  lubrication}}.  \jt{Phys. Rev. Lett.}  \bvol{92}~(24),  \pg{245509--1}.

\bibitem[Skotheim \& Mahadevan(2005)]{Skotheim2005a}
{\sc \au{Skotheim, J.~M.} \& \au{Mahadevan, L.}} \yr{2005}  \at{{Soft
  lubrication: The elastohydrodynamics of nonconforming and conforming
  contacts}}.  \jt{Phys. Fluids}  \bvol{17}~(9),  \pg{1--23}.

\bibitem[Snoeijer {\em et~al.\/}(2013)Snoeijer, Eggers \& Venner]{Snoeijer2013}
{\sc \au{Snoeijer, J.~H.}, \au{Eggers, J.} \& \au{Venner, C.~H.}} \yr{2013}
  \at{{Similarity theory of lubricated Hertzian contacts}}.  \jt{Phys. Fluids}
  \bvol{25}~(10).

\bibitem[Urzay {\em et~al.\/}(2007)Urzay, {Llewellyn Smith} \&
  Glover]{Urzay2007}
{\sc \au{Urzay, J.}, \au{{Llewellyn Smith}, S.~G.} \& \au{Glover, B.~J.}}
  \yr{2007}  \at{{The elastohydrodynamic force on a sphere near a soft wall}}.
  \jt{Phys. Fluids}  \bvol{19}~(10).

\bibitem[Venner \& Lubrecht(2000)]{venner2000multi}
{\sc \au{Venner, C.~H.} \& \au{Lubrecht, A.~A.}} \yr{2000} {\em Multi-Level
  Methods in Lubrication\/}. {\em Tribol. Interface Eng. Ser.\/} .
  \publ{Elsevier Science}.

\bibitem[Vialar {\em et~al.\/}(2019)Vialar, Merzeau, Giasson \&
  Drummond]{Vialar19}
{\sc \au{Vialar, P.}, \au{Merzeau, P.}, \au{Giasson, S.} \& \au{Drummond, C.}}
  \yr{2019}  \at{Compliant surfaces under shear: Elastohydrodynamic lift
  force}.  \jt{Langmuir}  \bvol{35}~(48),  \pg{15605--15613}.

\bibitem[Wang(2019)]{Wang2019}
{\sc \au{Wang, J.}} \yr{2019} {\em Interfacial mechanics: theories and methods
  for contact and lubrication\/}.  \publ{CRC Press}.

\bibitem[Wang {\em et~al.\/}(2015)Wang, Dhong \& Frechette]{wang2015}
{\sc \au{Wang, Y.}, \au{Dhong, C.} \& \au{Frechette, J.}} \yr{2015}
  \at{Out-of-contact elastohydrodynamic deformation due to lubrication forces}.
   \jt{Phys. Rev. Lett.}  \bvol{115},  \pg{248302}.

\bibitem[Zhang {\em et~al.\/}(2020)Zhang, Bertin, Arshad, Rapha\"el, Salez \&
  Maali]{Zhang20}
{\sc \au{Zhang, Z.}, \au{Bertin, V.}, \au{Arshad, M.}, \au{Rapha\"el, E.},
  \au{Salez, T.} \& \au{Maali, A.}} \yr{2020}  \at{Direct measurement of the
  elastohydrodynamic lift force at the nanoscale}.  \jt{Phys. Rev. Lett.}
  \bvol{124},  \pg{054502}.

\end{thebibliography}
\bibliographystyle{jfm}

\end{document}